\documentclass{osa-article}

\journal{osac}

\articletype{Research Article}

\begin{document}

\title{Maximizing optical production of metastable xenon}

\author{H.P. Lamsal, J.D. Franson, and T.B. Pittman}

\address{Physics Department, University of Maryland Baltimore
County, Baltimore, MD 21250}

\begin{abstract}
The wide range of applications using metastable noble gas atoms has led to a number of different approaches for producing large metastable state densities. Here we investigate a recently proposed hybrid approach that combines RF discharge techniques with optical pumping from an auxiliary state in xenon. We study the effect of xenon pressure on establishing initial population in both the auxiliary state and metastable state via the RF discharge, and the role of the optical pumping beam power in transferring population between the states. We find experimental conditions that maximize the effects, and provide a robust platform for producing relatively large long-term metastable state densities.
\end{abstract}

\section{Introduction}

Metastable noble gas atoms are used in diverse areas ranging from fundamental physics experiments \cite{giltner1995,helg1995, walhout1995, poupard2001}, to applications in plasma display panels \cite{tachibana2000,uhm2008,uhm2009}, ultralow-power nonlinear optical devices \cite{pittman2013,hickman2015,lamsal2019}, and rare gas lasers \cite{heaven2012,heaven2013,mikheyev2015,heaven2015,heaven2018,sanderson2019}.   In many of these applications, the production of high densities of metastable states is desirable, and techniques for efficiently promoting a substantial fraction of the ground state population to the metastable state are of paramount importance. Although relatively high metastable state densities ($\sim$$10^{13}$  cm$^{-3}$)  can be achieved for short time scales using pulsed electrical discharge techniques \cite{tachibana2000,uhm2008,uhm2009}, the realization of comparable long-term steady-state densities remains a significant challenge. Standard approaches include steady state RF excitation (see, for example, \cite{schearer1969,young2001,xia2010}), optical pumping techniques \cite{young2002,young2007,kohler2014}, and a recently proposed hybrid method which combines both \cite{hickman2016,zhang2020}.  In this paper, we investigate the experimental conditions needed to optimize this hybrid approach to maximize the production of metastable state densities.  Our study specifically investigates the production of metastable xenon (Xe*), but the general results are expected to apply to other noble gas species as well.

The energy level diagrams in Figure 1 provide a simplified comparison of the various methods used to promote the ground state, $|g\rangle$, of a noble gas atom to the metastable state of interest, $|m\rangle$. Here, $|m\rangle$ lies roughly $\sim$$10$ eV above $|g\rangle$, and optical transitions between $|g\rangle$ and $|m\rangle$ are forbidden by electric dipole selection rules.  Noble gas atoms also have a nearby auxiliary state, $|a\rangle$, which does have a direct optical transition from $|g\rangle$ at a resonance wavelength in the deep UV ( $\sim$$100$ nm). Finally, we consider an upper-level state, $|u\rangle$, which has NIR optical transitions ($\sim$800 nm ) between both $|a\rangle$ and $|m\rangle$.

\begin{figure}[h]
\centering\includegraphics[width=5.2in]{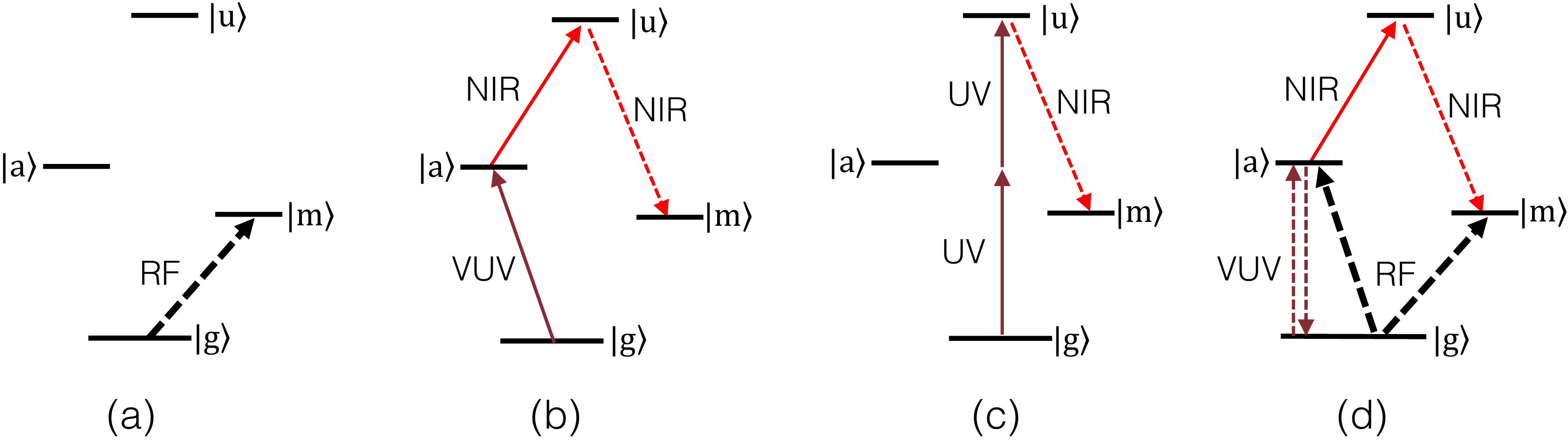}
\caption{Comparative overview of 4 different methods used to promote ground state $|g\rangle$ population to the metastable state $|m\rangle$ in noble gas atoms: (a) standard RF discharge technique, (b) ``all-optical'' technique involving transitions to an auxiliary state $|a\rangle$ and upper state $|u\rangle$ \cite{young2002,young2007,kohler2014}, (c) alternative ``all-optical'' technique \cite{gornik1981,dakka2018}, and (d) the hybrid technique of interest here \protect\cite{hickman2016}. The hybrid technique in panel (d) combines the electrical discharge of panel (a) with the optical pumping of panel (b).}
\end{figure}

Panel (a) of Figure 1 represents a traditional electrical discharge technique  \cite{schearer1969,young2001,xia2010}. Here an RF (or DC) field is applied to a gas of atoms, resulting in various collisional mechanisms that transfer population (both directly and indirectly) from $|g\rangle$ to $|m\rangle$.  In contrast, panel (b) shows an ``all optical'' approach in which resonant VUV light (typically from an external lamp) excites $|g\rangle \rightarrow |a\rangle$ \cite{young2002,young2007,kohler2014}.  Population is then optically pumped from  $|a\rangle \rightarrow |u\rangle$ using a narrowband laser, and spontaneous emission then causes a transition from  $|u\rangle \rightarrow |m\rangle$.  Panel (c) shows an alternative ``all-optical'' approach that relies on a direct two-photon transition from  $|g\rangle \rightarrow |u\rangle$ \cite{gornik1981,dakka2018}.

Panel (d) of Figure 1 shows the hybrid approach of interest here \cite{hickman2016}.  The key idea is that the same electrical discharge that creates population in $|m\rangle$ (cf. panel (a)) also creates population in the auxiliary state $|a\rangle$. This electrically-produced auxiliary population can then be optically pumped from  $|a\rangle \rightarrow |u\rangle \rightarrow |m\rangle$ to supplement the existing electrically-produced $|m\rangle$ population.  At the same time, spontaneous emission from the dipole-allowed $|a\rangle \rightarrow |g\rangle$ transition provides a source of resonant VUV photons that can be subsequently absorbed by nearby ground state atoms. Because the overall number of atoms that are excited by the RF discharge in a typical experimental setup is much larger than the number of atoms in the cross-section of the NIR optical pumping beam, this effect gives a significant boost to the population that can be optical pumped from $|a\rangle \rightarrow |m\rangle$.   The combination of all of these effects results in the strong optical enhancement to the electrically-produced population of $|m\rangle$ that was first observed in \cite{hickman2016}.

In this paper we investigate the experimental conditions that maximize this optically-enhanced electrical production of metastable states. We use a strong RF discharge and spectroscopic techniques to study and compare the initial populations of both the metastable state $|m\rangle$ and the auxiliary state $|a\rangle$, and find that they are each maximized under the same experimental conditions.  In addition, we experimentally observe that the metastable state population can be enhanced by an amount greater than the steady-state population of the auxiliary state due to the drastically different effective lifetimes of these states.  The end result is a convenient and robust method to experimentally optimize the overall steady-state metastable state population using the hybrid technique illustrated in Figure 1(d).

\section{Metastable xenon}

Our study uses a low pressure ($\sim$100 mTorr) gas of xenon atoms. An overview of the relevant energy levels is shown in Figure 2.  Here, the metastable state $|m\rangle$ ($6s[3/2]_{2}$ state)  has a natural lifetime of $\sim$43 s\cite{walhout1994} which is reduced to the order of ms due to collisions under our typical experimental operating conditions \cite{molnar1953,futch1956,barbet1975}.  The auxiliary state $|a\rangle$ ($6s[3/2]_{1}$ state) has a lifetime of $\sim$4 ns, which is significantly shorter than the lifetime of $|m\rangle$. In xenon, the UV optical transition between $|a\rangle$ and the ground state $|g\rangle$ ($5P^{6}$ state) is at 147 nm, and the NIR transition between $|a\rangle$ and the upper state $|u\rangle$ ( $6p[3/2]_{1}$ state) is at 916 nm. This 916 nm transition is driven by a narrowband pump laser in our experiments. 

\begin{figure}[h]
\centering\includegraphics[width=3.9in]{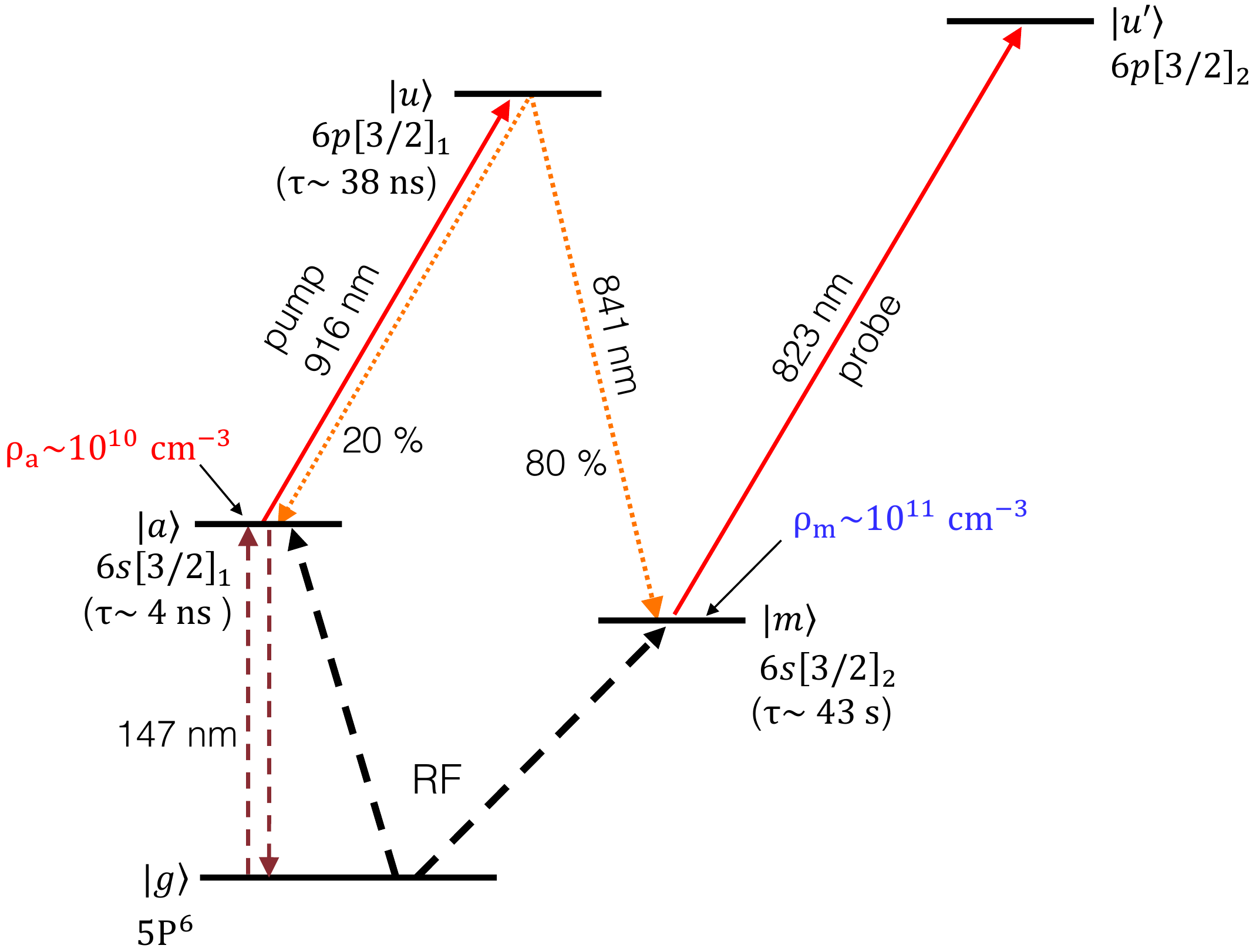}
\caption{Relevant energy level diagram and associated parameters for xenon. Here, the metastable state $|m\rangle$ has an intrinsic lifetime of $\tau$$\sim$43 sec, while the auxiliary state $|a\rangle$ has a lifetime of $\tau$$\sim$4 ns. In our experiments, the baseline densities (due to the RF discharge alone) of the metastable state and auxiliary state are $\rho_{m}$$\sim$$10^{11}$cm$^{-3}$ and $\rho_{a}$$\sim$$10^{10}$ cm$^{-3}$, respectively. The goal of the experiments is to optically pump as much of the $|a\rangle$ population to  $|m\rangle$ as possible. The metastable state density $\rho_{m}$ is measured by performing spectroscopy on an 823 nm transition between $|m\rangle$ and a second upper-level state $|u'\rangle$. }
\end{figure}

Once populated, $|u\rangle$  can decay to the desired metastable state $|m\rangle$ via spontaneous emission at 841 nm with a probability of 80\%,  or back to the auxiliary state $|a\rangle$ with a probability of 20\%. The lifetime of the upper state $|u\rangle$ is  $\sim$38 ns \cite{alford1992}.   In addition to the four key atomic states of interest, we also utilize an extra upper state $|u'\rangle$ ( $6p[3/2]_{2}$ state) to perform spectroscopic measurements that characterize the metastable state (Xe*) density as a function of various experimental parameters.

As will be described below, the baseline steady-state Xe* density under optimized RF discharge conditions in our experiment is $\sim$$10^{11}$ cm$^{-3}$.   Under the same conditions, the baseline density of atoms in the auxiliary state $|a\rangle$ is  $\sim$$10^{10}$ cm$^{-3}$. The goal of optically enhanced production of Xe* is simply to transfer as much of this $|a\rangle$ population as possible into $|m\rangle$ \cite{hickman2016}.  Intuitively, the requirements are (1) starting with a large population in $|a\rangle$, and (2) driving the process with a strong optical pumping beam at 916 nm.  In the following sections, we summarize a series of experimental measurements that characterize the optimization of this process. The key parameters that influence the process are the overall pressure of the neutral xenon gas, and the strength of the 916 nm optical pumping beam.

\section{Experimental setup}

Figure 3 provides an overview of the pump-probe experimental setup. The pump and probe lasers are independent narrowband tunable diode lasers (Toptica DL Pro; linewidths $\sim$300 KHz) centered at 916 nm and 823 nm, respectively.  For the spectroscopic measurements of interest, the lasers are scanned over ranges of $\sim$10 GHz, and the laser frequencies are measured using a calibrated wavelength meter with 100 MHz resolution (High Finesse WSU-30). The pump and probe lasers are combined into a single spatial mode using a single-mode fiber, and then launched into the xenon vacuum chamber as a common free-space beam with a diameter of $\sim$2 mm. A power meter that can be inserted into the free-space beam before the vacuum chamber is used to measure the laser powers.  To avoid atomic saturation effects, the experiments are done with 823 nm probe beam powers of only $\sim$7.5 uW \cite{jones2014}. As will be described below, the 916 nm pump beam power is a key experimental parameter that is varied in the range of 0 - 10 mW.

\begin{figure}[t]
\centering\includegraphics[width=5.24in]{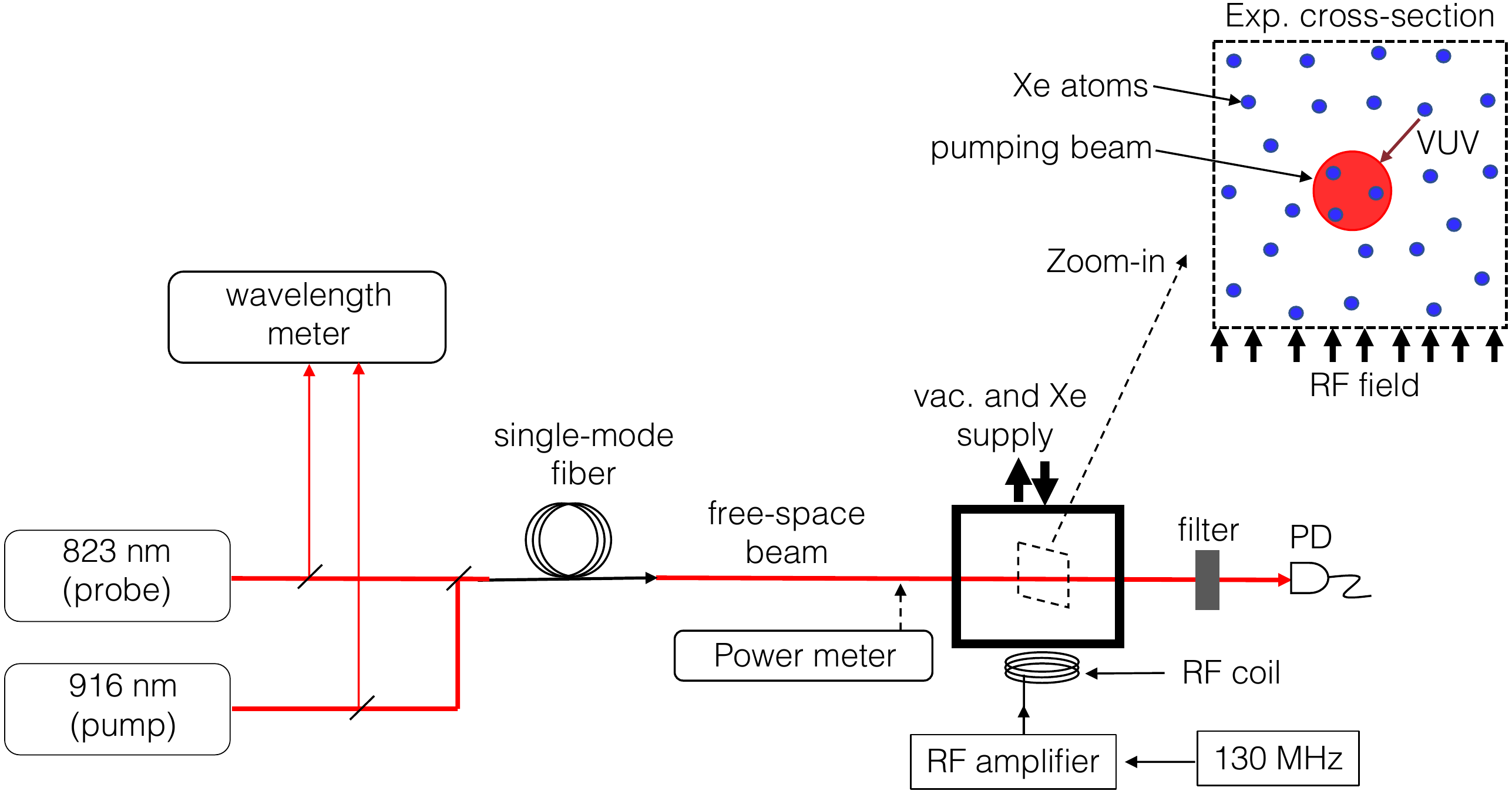}
\caption{Overview of the pump-probe setup used to characterize the experimental conditions that maximize the optical production of Xe*. The pump and probe lasers (at 916 nm and 823 nm, respectively) are combined into a common spatial mode using a single-mode fiber, and then launched as a free-space beam through a vacuum chamber containing a gas of xenon atoms.  A xenon glow discharge plasma is produced using an RF coil mounted on the outside of a glass window on the vacuum chamber. The dashed-box inset represents a ``zoom-in'' on a cross-section of the experiment. The RF field excites atoms throughout the entire chamber, while the 2mm diameter pumping beam (and probe beam) only interact with a small subset of the atoms. The 2 key experimental parameters are the neutral xenon pressure and the 916 nm pumping beam power.}
\end{figure}

The vacuum system is comprised of a standard 4.5" stainless steel ConFlat cube with glass viewports that let the pump and probe beams pass through the chamber.  The chamber is typically pumped down to a pressure of $\sim$$10^{-6}$ Torr and then backfilled with xenon gas. A capacitance manometer gauge fitted on the chamber is used to monitor the neutral xenon pressure.  The neutral xenon pressure is a key experimental parameter that it is varied the range of $\sim$10 - 600 mTorr in the experiments. 

A large 4.5" glass viewport is also used on one side of the cube to allow RF excitation of the xenon gas.  An RF coil is mounted on the viewport, and driven with a standard tank circuit (consisting of the coil and capacitors).  This system has an RF resonance frequency of 130 MHz, and is driven by a low power signal generator that is amplified to produce a glow discharge plasma in the chamber. For our full range of neutral xenon pressures, an amplified RF power of  $\sim$20 W was found to maximize the initial Xe* density \cite{lamsal2019} (ie. the baseline $|m\rangle$ population; via the process shown illustrated in Figure 1(a)).

The inset to Figure 3 illustrates a cross-section of the gas of atoms under the influence of both the RF excitation field, and the optical pumping beam at 916 nm. This highlights the fact that while a large $\sim$10 cm diameter cross section undergoes RF excitation, only a small subset of the atoms interact with the optical pumping beam ($\sim$2 mm diameter). These surrounding atoms provide a source of spontaneously emitted 147 nm photons that boost the baseline population of $|a\rangle$ in Figure 2. 

A photodiode (PD) preceded by various filters is used to measure the transmitted laser powers during the spectroscopic measurements.  For the baseline spectroscopic measurements of the $|a\rangle$ to $|u\rangle$ transition at 916 nm, or the $|m\rangle$ to $|u'\rangle$ transition at 823 nm, only a single laser is used.  For the full pump-probe experiments using both lasers, a narrow bandpass filter (centered at 823 nm) is used to block the transmitted 916 nm pump beam, while passing the 823 nm probe beam during the measurements.

\section{Experimental results}

With the absence of the RF field, nearly all of the atoms in the gas are in the ground state and the initial populations of $|a\rangle$ and $|m\rangle$ are roughly zero. The RF discharge contribution to the population of $|m\rangle$ can then be determined by applying the RF field, and performing transmission-spectroscopy measurements on the $|m\rangle \rightarrow |u'\rangle$ transition at 823 nm. An example result at a neutral xenon pressure of 15 mTorr  is shown in Figure 4(a).  Here, the 6 ``dips'' in transmission are due to absorption from the various isotopes of Xe \cite{xia2010}. We obtain an estimate of the Xe* density to be $\rho_{m}$$\sim$2 x $10^{11}$ cm$^{-3}$ by fitting these transmission lineshapes using the models of reference \cite{lin2001}, while taking into account the natural abundances of the various xenon isotopes and branching ratios of the hyperfine transitions \cite{agnes2013}. This value of  $\rho_{m}$$\sim$2 x $10^{11}$ cm$^{-3}$ represents the baseline steady-state Xe* density (ie. $|m\rangle$ population) that can subsequently be enhanced by optical pumping population from the auxiliary state $|a\rangle$.

\begin{figure}[b]
\centering\includegraphics[width=4.8in]{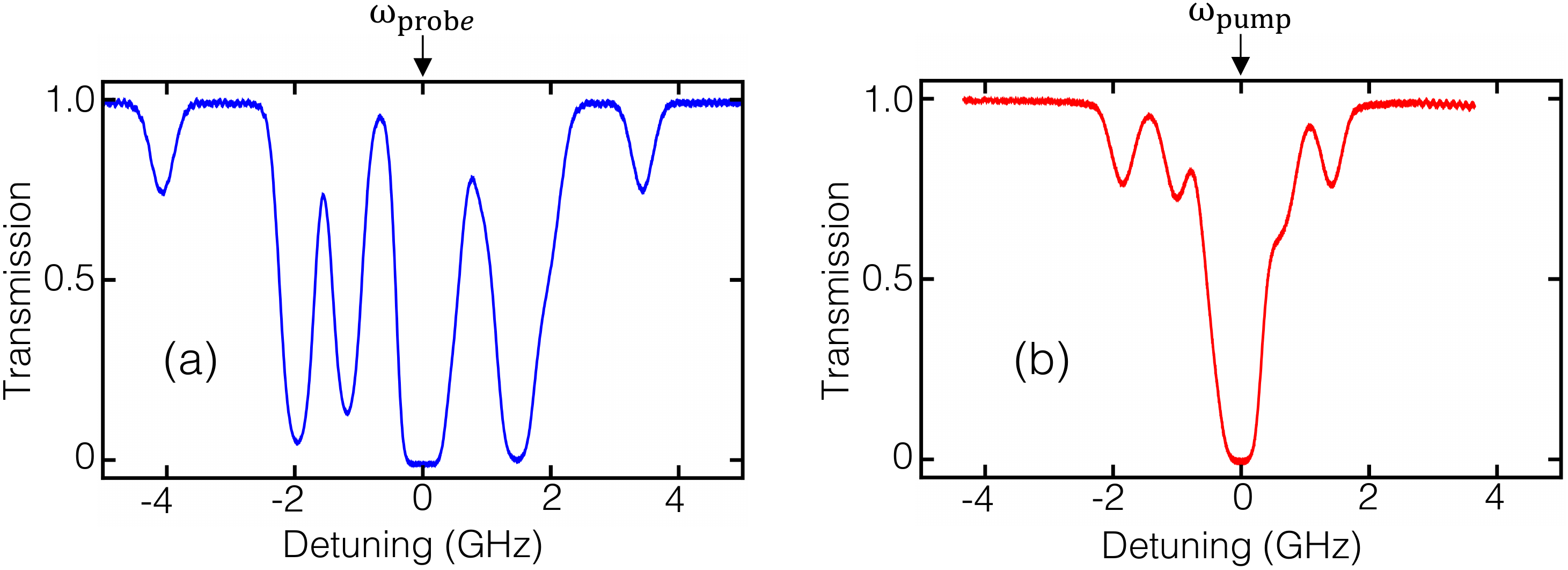}
\caption{Single laser spectroscopy of (a) the $|m\rangle \rightarrow |u'\rangle$ transition at 823 nm, and (b) the $|a\rangle \rightarrow |u\rangle$ transition at 916 nm. In panel (a) a detuning of zero corresponds to $\omega_{probe}$ = 364,095 GHz. In panel (b), zero detuning corresponds to $\omega_{pump}$ = 327,099 GHz.  Throughout the work, transmission spectra lineshapes like these are fitted to determine the metastable state $|m\rangle$ and auxiliary state $|a\rangle $ densities $\rho_{m}$ and $\rho_{a}$.}
\end{figure}

We estimate the baseline population of $|a\rangle$ by performing analogous transmission-spectroscopy measurements on the $|a\rangle \rightarrow |u\rangle$ transition at 916 nm. Figure 4(b) shows a typical result using the same RF discharge conditions and xenon pressure of 15 mTorr. Here, the large central ``dip'' and smaller side dips are once again due to the various isotopes of Xe and the hyperfine splittings of the relevant states \cite{kramida2019}. Fitting the lineshapes in Figure 4(b) provides an estimated baseline $|a\rangle$ density of $\rho_{a}$$\sim$4 x $10^{10}$ cm$^{-3}$.  Note that this population of $|a\rangle$ results from both the RF discharge, as well as the absorption of 147 nm photons spontaneously emitted from surrounding atoms in the large-scale gas.

Next, we repeat these independent measurements for a range of neutral xenon pressures ranging from 10 to 600 mTorr. The results are shown in Figure 5, where each data point represents a baseline density value extracted from fitting a transmission spectrum similar to those in Figure 4.  The key result here is the nearly identical behavior of the $|m\rangle$ and $|a\rangle$ populations as function of the xenon pressure in our system. As the pressure is increased there are simply more atoms to promote from the ground state and the populations increase, while at higher pressures the role of collisions plays a more dominant role and limits the populations. Note that the trade-off between these effects results in the same optimal pressure of 15 mTorr  for both the $|m\rangle$ and $|a\rangle$ populations in our system. At pressures higher than $\sim$600 mTorr (or lower than 10 mTorr), we were unable to produce a stable discharge in our system.

From a practical point of view, the significance of the nearly identical behavior in Figure 5 is that the same experimental conditions that give the largest baseline Xe* density also provide the largest auxiliary state density. Because the optical enhancement of Xe* production intuitively requires a large baseline auxiliary state population, this result suggests that the conditions needed for maximizing the optically-enhanced production of Xe* can be achieved by simply maximizing the starting baseline density of Xe*.

\begin{figure}[t]
\centering\includegraphics[width=3.4in]{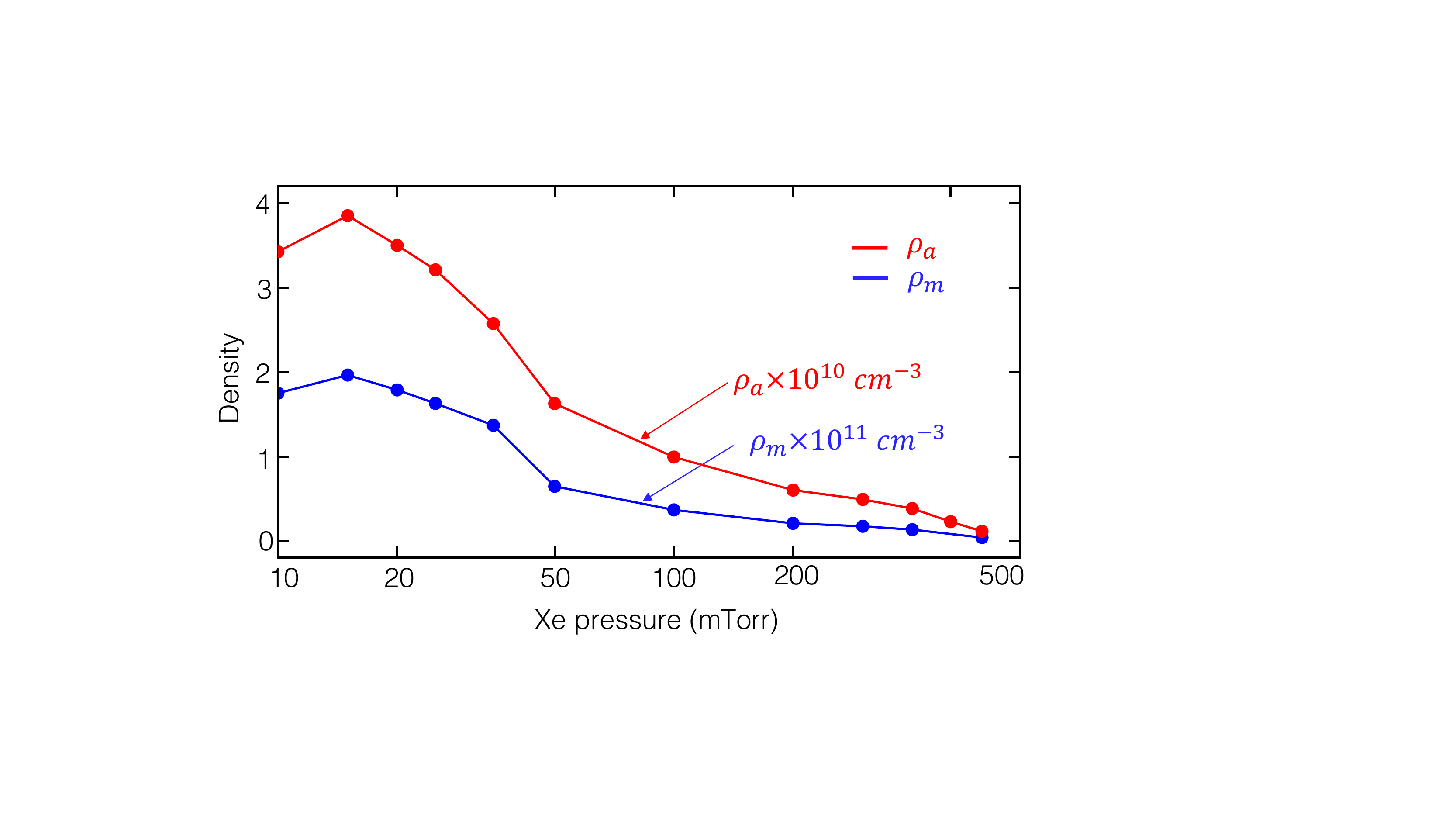}
\caption{Baseline densities (due to the RF discharge alone) of the metastable state $|m\rangle$ (blue trace) and auxiliary state $|a\rangle$ (red trace) as a function of neutral xenon pressure. The key result here is the similar dependence on pressure; optimizing the system for the largest baseline $\rho_{m}$ value also provides the largest baseline $\rho_{a}$ value.}
\end{figure}

\begin{figure}[b]
\centering\includegraphics[width=5.3in]{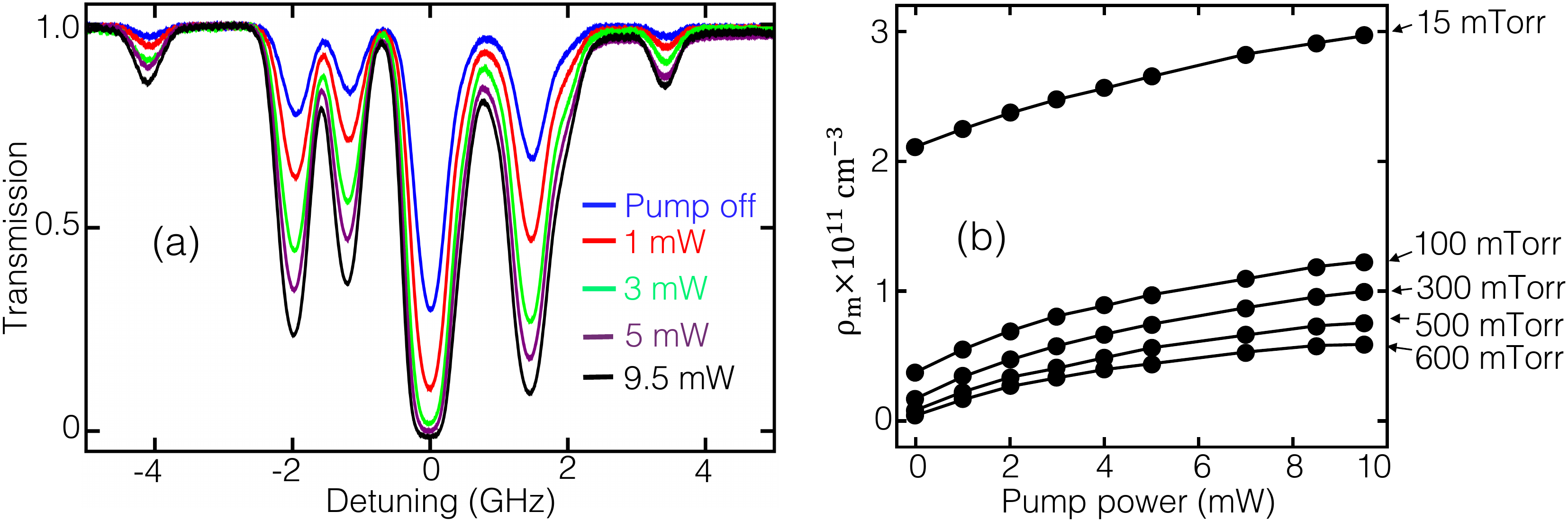}
\caption{(a) example of full pump-probe experimental results at a neutral xenon pressure of 300 mTorr. The data shows 823 nm probe spectroscopy for several different values of 916 nm pump beam power $P_{pump}$ ranging from 0 mW (pump off) to 9.5 mW.  As $P_{pump}$ is increased, the increasing absorption ``dips''  are indicative of more population being pumped from $|a\rangle$ to $|m\rangle$. (b) quantification of the desired increase in $\rho_{m}$ as a function of  $P_{pump}$ for several different neutral xenon pressures in the range of 15 mTorr to 600 mTorr.}
\end{figure}

In Figure 6 we investigate the role of the 916 nm optical pumping beam power, $P_{pump}$, for transferring population from $|a\rangle$ to $|m\rangle$. We perform a series of pump-probe experiments for a variety of neutral xenon pressures.  Here,  823 nm (probe) transmission spectra similar to those in Figure 4(a) are measured, but now with the 916 nm optical pumping beam turned on (and with $\omega_{pump}$ fixed at zero detuning). Figure 6(a) shows an example of results obtained at a relatively high neutral xenon pressure of 300 mTorr. As $P_{pump}$ is increased, more population is transferred from $|a\rangle$ to $|m\rangle$, resulting in stronger absorption of the 823 nm probe beam (ie. increasingly deeper ``dips'' in the transmission spectra).

In order to quantify this population transfer, we fit each of the spectra in Figure 6(a) to obtain estimates of the Xe* density $\rho_{m}$ as a function of $P_{pump}$. We then repeat these pump-probe measurements for a several different neutral xenon pressures ranging from 15 to 600 mTorr, with the results summarized in Figure 6(b).  In all cases, a dramatic increase of  $\rho_{m}$ is seen as a function of increasing $P_{pump}$.  In many ways, the results shown in Figure 6(b) represent the main result of the study. We see that the goal of maximizing the metastable state density $\rho_{m}$ is achieved by  using the neutral xenon pressure that maximizes the initial baseline values of $\rho_{m}$ (and thus $\rho_{a}$), and utilizing as much pump beam power as possible. A saturation-like effect is also seen for all cases of neutral xenon pressure \cite{hickman2016,zhang2020}.

The overall utility of this hybrid approach is seen most clearly for non-optimized neutral xenon pressures which result in very small initial baseline densities $\rho_{m}$ (for example, the 600 mTorr case in Figure 6(b)). Here, the application of pump beam powers of only a few mW can provide a relative increase in $\rho_{m}$ by roughly an order of magnitude \cite{hickman2016}.  However, one significant result which emerges from the data in Figure 6(b) is that these order-of-magnitude increases do not hold for more optimized cases of larger initial baseline densities (for example, the 15 mTorr case in Figure 6(b)).  Rather, the data in Figure 6(b) show that the {\em net} increase in $\rho_{m}$ as function of $P_{pump}$ is roughly constant over a large range of neutral xenon pressures.  Figure 7 examines this effect in more detail by plotting the net increase in metastable state density,  $\Delta \rho_{m}$, as a function of neutral xenon pressure, for several different values of $P_{pump}$.  The data shows that over the range of xenon pressures from 15 mTorr  to $\sim$300 mTorr,  $\Delta \rho_{m}$ is essentially flat, with a value determined solely by $P_{pump}$. 

\begin{figure}[b]
\centering\includegraphics[width=4.4in]{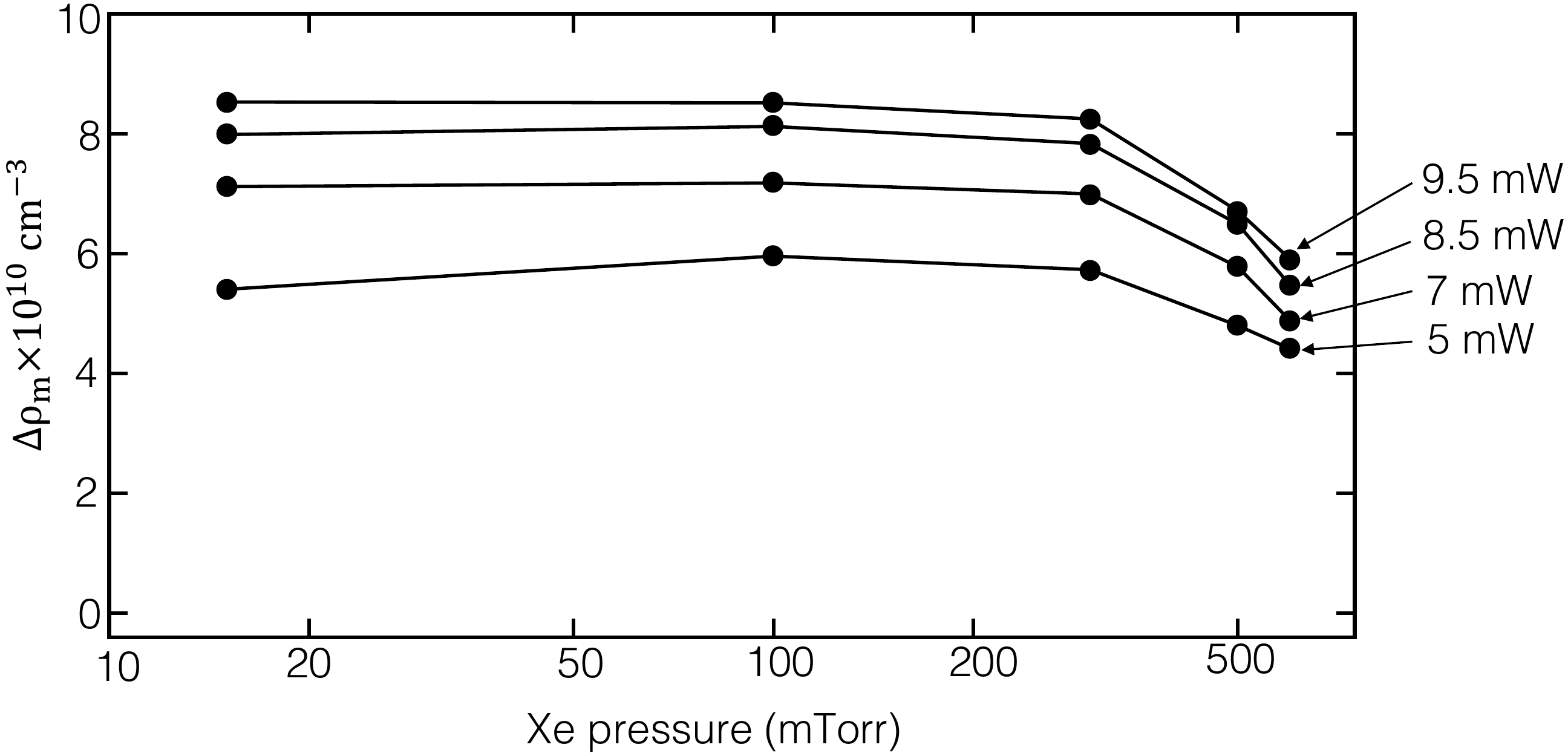}
\caption{Net change in metastable state density as a function of neutral xenon pressure for 4 different values of pump beam power between a moderate value of 5 mW and our maximum available power of 9.5 mW.  The data shows that $\Delta \rho_{m}$ is fairly constant over a large pressure range, with a value largely determined by the pump power.}
\end{figure}

It is interesting to note that the value of $\Delta\rho_{m}$ (for all cases in Figure 7) actually {\em exceeds} the initial steady-state baseline auxiliary state density $\rho_{a}$.  For example, using our maximum available pump power of $P_{pump}$ = 9.5 mW (upper trace in Figure 7), the change in Xe* density is roughly $\Delta \rho_{m}$$\sim$8.5 x $10^{10} $cm$^{-3}$ over the range of xenon pressures from 15 mTorr  to $\sim$300 mTorr.  Over that same pressure range, the baseline auxiliary state density $\rho_{a}$ varies from a maximum of only 3.9 x $10^{10}$ cm$^{-3}$ down to 0.1 x  $10^{10}$ cm$^{-3}$ ({\em cf.} Figure 5).  This apparent excess in  $\Delta \rho_{m}$ is due to the fact that the metastable state effective lifetime ($\sim$ms) is several orders of magnitude larger than the auxiliary state $|a\rangle$ and upper state $|u\rangle$ lifetimes ($\sim$ns), resulting in a desirable continuous-wave laser-pumped steady-state ``build up'' of population in the metastable state \cite{happer1972}.   In Figure 7 the small drop-off in $\Delta \rho_{m}$ values for pressures higher than $\sim$300 mTorr is most-likely due to insufficient initial $\rho_{a}$ values in our set-up at higher pressures ({\em cf.} Figure 5).

\section{Summary and conclusions}

We have investigated the experimental conditions needed to maximize the optical production of metastable xenon using the hybrid method of Figure 1(d) \cite{hickman2016,zhang2020}.  Here, conventional RF-excitation is combined with optical pumping from an auxiliary state to provide a dramatic enhancement of the metastable state density. We find that the same RF discharge conditions that optimize the initial metastable state density also optimize the auxiliary state density. While these initial baseline densities (due to RF excitation alone) vary significantly with the neutral xenon pressure, we find that the subsequent net increase in metastable state density (due to optical pumping from the auxiliary state) is relatively constant over a large pressure range, and is determined solely by the pumping beam power. In addition, the net increase in the steady-state metastable state density can exceed the baseline auxiliary state density due to the vastly different lifetimes of these states.  

These results suggest that the hybrid technique of Figure 1(d) can provide a robust method for producing relatively large long-term metastable state densities. It is important to note, however, that the long-term densities achieved here ($\sim$$10^{11}$ cm$^{-3}$) are still lower than those obtained using short pulsed electrical discharge techniques ($\sim$$10^{13}$ cm$^{-3}$) \cite{tachibana2000,uhm2008,uhm2009}. Nonetheless, the techniques described here impact applications requiring high densities for longer continuous time scales \cite{sanderson2019}, and studying the transient dynamics of this hybrid approach in the pulsed regime may also be of fundamental interest.

\section*{Funding}
National Science Foundation (Grant No. 1402708); The Office of Naval Research (Grant N00014-15-1-2229)


\begin{thebibliography}{50}

\bibitem{giltner1995} David M. Giltner, Roger W. McGowan, and Siu Au Lee, ``Atom Interferometer Based on Bragg Scattering from Standing Light Waves'', Phys. Rev. Lett. \textbf{75}, 2638 (1995).

\bibitem{helg1995} K.K. Berggren, A. Bard, J.L. Wilbur, J.D. Gillaspy, A.G. Helg, J.J. McClelland, S.L. Rolston, W.D. Phillips, M. Prentiss, and G.M. Whitesides, ``Microlithography by using neutral metastable atoms and self-assembled monolayers'', Science, \textbf {269}, 1255 (1995).

\bibitem{walhout1995} M. Walhout, U. Sterr, C. Orzel, M. Hoogerland, and S.L. Rolston, ``Optical Control of Ultracold Collisions in Metastable Xenon'', Phys. Rev. Lett. \textbf{74}, 506 (1995).

\bibitem{poupard2001} A. Robert, O. Sirjean, A. Browaeys, J. Poupard, S. Nowak, D. Boiron, C.I. Westbrook, and A. Aspect, ``A Bose-Einstein Condensate of Metastable Atoms'', Science \textbf{292}, 461 (2001). 

\bibitem{tachibana2000} K. Tachibana, and S. Feng, and T. Sakai, ``Spatiotemporal behaviors of excited Xe atoms in unit discharge cell of ac-type plasma display panel studied by laser spectroscopic microscopy'', J. Appl. Phys. \textbf{88},  4967 (2000).

\bibitem{uhm2008} H.S. Uhm, P.Y. Oh, and E.H. Choi, ``Properties of excited xenon atoms in an alternating current plasma display panel'', Appl. Phys. Lett. \textbf{93}, 211501 (2008). 

\bibitem{uhm2009} H.S. Uhm, B.H. Hong, P.Y. Oh, E.H. Choi, ``Properties of excited xenon atoms in a plasma display panel'', Thin Solid Films \textbf{517}, 4023 (2009).

\bibitem{pittman2013} T.B. Pittman, D.E. Jones, and J.D. Franson, ``Ultralow-power nonlinear optics using tapered optical fibers in metastable xenon'', Phys. Rev. A \textbf{88}, 053804 (2013). 

\bibitem{hickman2015} G.T. Hickman, T.B. Pittman, and J.D. Franson, ``Low-power cross-phase modulation in a metastable xenon-filled cavity for quantum-information applications'', Phys. Rev. A \textbf{92}, 053808 (2015).

\bibitem{lamsal2019} H.P. Lamsal, J.D. Franson, and T.B. Pittman, ``Transmission characteristics of optical nanofibers in metastable xenon'',  Appl. Opt.  \textbf{58}, 6470 (2019).

\bibitem{heaven2012} J. Han and M.C. Heaven, ``Gain and lasing of optically pumped metastable rare gas atoms'', Opt. Lett. \textbf{37}, 2157 (2012).

\bibitem{heaven2013} J. Han, L. Glebov, G. Venus, and M.C. Heaven, ``Demonstration of a diode-pumped metastable Ar laser'', Opt. Lett. \textbf{38}, 5458 (2013).

\bibitem{mikheyev2015} P. A. Mikheyev, ``Optically pumped rare-gas lasers'', Quantum. Electron. \textbf{45}, 704 (2015).

\bibitem{heaven2015} W.T. Rawlins, K.L. Galbally-Kinney, S.J. Davis, A.R. Hoskinson, J.A. Hopwood, and M.C. Heaven, ``Optically pumped microplasma rare gas laser'', Opt. Express \textbf{23}, 4804 (2015).

\bibitem{heaven2018} A.V. Demyanov, I.V. Kochetov, P.A. Mikheyev, V.N. Azyazov and M.C. Heaven, ``Kinetic analysis of rare gas metastable production and optically pumped Xe lasers'', Journal of Physics D: Applied Physics, \textbf{51}, 045201 (2018).

\bibitem{sanderson2019} C.R. Sanderson, C.W. Ballmann, J. Han, A.B. Clark, B.H Hokr, K.G. Xu, and M.C. Heaven, ``Demonstration of a quasi-CW diode-pumped metastable xenon laser'', Opt. Express \textbf{27}, 36011 (2019).

\bibitem{schearer1969} L.D. Schearer, ``Optical Pumping of Neon $^3$P$_2$ Metastable Atoms'', Phys. Rev. \textbf{180}, 83 (1969).

\bibitem{young2001} C.Y. Chen, K. Bailey, Y.M. Li, T.P. O’Connor, Z.-T. Lu, X. Du, L. Young and G. Winkler, ``Beam of metastable krypton atoms extracted from a rf-driven discharge'', Rev. Sci. Instrum. \textbf{72}, 271 (2001). 

\bibitem{xia2010} T. Xia, S.W. Morgan, Y.-Y. Jau, and W. Happer, ``Polarization and hyperfine transitions of metastable $^{129}$Xe in discharge cells'', Phys. Rev. A \textbf{81}, 033419 (2010). 

\bibitem{young2002} L. Young and D. Yang and R.W. Dunford,  ``Optical production of metastable krypton'',  J. Phys. B: At. Mol. Opt. Phys. \textbf{35}, 2985 (2002). 

\bibitem{young2007}Y. Ding, S.-M. Hu, K. Bailey, A.M. Davis, R.W. Dunford, Z.-T. Lu, T.P. O’Connor, and L. Young, ``Thermal beam of metastable krypton atoms produced by optical excitation'', Rev. Sci. Instrum. \textbf{78}, 023103 (2007).

\bibitem{kohler2014} M. Kohler, H. Daerr, P. Sahling, C. Sieveke, N. Jerschabek, M.B. Kalinowski, C. Becker, and K. Sengstock,  ``All-optical production and trapping of metastable noble-gas atoms down to the single-atom regime'', Europhys. Lett. \textbf{108}, 13001 (2014).

\bibitem{hickman2016} G.T. Hickman, J.D. Franson, and T.B. Pittman, ``Optically enhanced production of metastable xenon'', Opt. Lett. \textbf{41}, 4372 (2016).

\bibitem{zhang2020} Z.-Y. Zhang, F. Ritterbusch, W.-K. Hu, X.-Z. Dong, C.Y. Gao, W. Jiang, S.-Y. Liu,
Z.-T. Lu, J.S. Wang, and G.-M. Yang, ``Enhancement of the $^{81}$Kr and $^{85}$Kr count rates by optical pumping'', Phys. Rev. A \textbf{101}, 053429 (2020).

\bibitem{gornik1981} W. Gornik, S. Kindt, E. Matthias, and D. Schmidt, ``Two-photon excitation of xenon atoms and dimers in the energy region of the 5p56p configuration'', J. Chem. Phys. \textbf{75}, 68 (1981).


\bibitem{dakka2018} M.A. Dakka, G. Tsiminis, R.D. Glover, C. Perrella, J. Moffatt, N.A. Spooner, R.T. Sang, P.S. Light, and A.N. Luiten, ``Laser-Based Metastable Krypton Generation'', Phys. Rev. Lett. \textbf{121}, 093201 (2018).

\bibitem{walhout1994}  M. Walhout, A. Witte, and S.L. Rolston, ``Precision measurement of the metastable 6s[3/2]$_2$  lifetime in xenon'', Phys. Rev. Lett. {\bf 72}, 2843 (1994).

\bibitem{molnar1953} A.V. Phelps and J.P. Molnar, ``Lifetimes of Metastable States of Noble Gases'', Phys. Rev.  \textbf{89}, 1202 (1953).

\bibitem{futch1956} A.H. Futch and F.A.Grant, ``Mean Life of the $^3$P$_2$ Metastable Argon Level'', Phys. Rev.  \textbf{104}, 356 (1956).

\bibitem{barbet1975} A. Barbet, N. Sadeghi and J.C. Pebay-Peyroula, ``Decay of metastable xenon atoms Xe*($^3$P$_2$) in a xenon afterglow'', J. Phys. B: At. Mol. Phys. \textbf{8}, 1776 (1975).

\bibitem{alford1992} W.J. Alford, ``State-to-state rate constants for quenching of xenon 6p levels by rare gases'', J. Chem. Phys. \textbf{96}, 4330 (1992).

\bibitem{jones2014} D.E. Jones, J.D. Franson, and T.B. Pittman, ``Saturation of atomic transitions using sub-wavelength diameter tapered optical fibers in rubidium vapor'' J. Opt. Soc. Am. B \textbf{31}, 1997 (2014).

\bibitem{lin2001} Y. Liu, J. Lin, G. Huang, Y. Guo, and C. Duan, ``Simple empirical analytical approximation to the Voigt profile'', J. Opt. Soc. Am. B {\bf 18}, 666 (2001).

\bibitem{agnes2013} P. Jacquet and A. Pailloux, ``Laser absorption spectroscopy for xenon monitoring in the cover gas of sodium cooled fast reactors'', J. Anal. At. Spectrom. {\bf 28}, 1298 (2013). 
\bibitem{kramida2019}A. Kramida, Yu. Ralchenko, J. Reader, and NIST ASD Team, Atomic Spectra Database (National Institute of Standards and Technology, Gaithersburg, MD, 2019) [https://physics.nist.gov/asd].

\bibitem{happer1972} W. Happer, ``Optical Pumping", Rev. Mod. Phys. \textbf{44}, 169 (1972).


\end{thebibliography}
\end{document}